\documentclass[twocolumn,pra,showpacs,superscriptaddress,amssymb,amsmath,amsmath,floatfix]{revtex4-1}
\usepackage{graphicx}
\usepackage{epstopdf}
\usepackage{bm}
\usepackage[colorlinks=true,linkcolor=blue,urlcolor=blue,citecolor=blue]{hyperref}
\usepackage{comment}
\usepackage{float}
\usepackage{cancel}
\usepackage{times}
\usepackage{amsmath, amsthm, amssymb, amsfonts}
\usepackage{xcolor}

\newcommand{\hl}[1]{\textcolor{black}{#1}}
\newcommand{\kj}[1]{\textcolor{black}{#1}}
\newcommand{\kb}[1]{\textcolor{black}{#1}}
\newcommand{\kr}[1]{\textcolor{black}{#1}}

\newcommand{\kn}[1]{\textcolor{black}{#1}}

\begin{document}
\author{Krzysztof Myśliwy and Marek Napiórkowski
\\ \emph{Institute of Theoretical Physics, Faculty of Physics, University of Warsaw \\ Pasteura 5, 02-093 Warsaw, Poland}}
\title{Fully polarized Fermi systems at finite temperature}

\begin{abstract}
We propose a simple model of an interacting, fully spin--polarized Fermi gas in dimensions $d=2$ and $d=3$, and derive the approximate expression for the energy spectrum and the corresponding formula for the Helmholtz free energy. We analyze the thermodynamics of the system and find the lines of first--order phase transitions between the low and high density phases terminating at critical points. The properties of the corresponding phase diagrams are qualitatively different for $d=2$ and $3$, and sensitively depend on the interparticle attraction, which marks a departure from the standard van der Waals theory. The differences originate from the Pauli exclusion principle and are embeded in the fermionic nature of the system under study.  
  \end{abstract}
\maketitle

\section{Introduction}

The properties of matter are largely determined by the mutual interactions between the constituent particles. If, however,  the system in question is composed of fermions the Pauli exclusion principle \cite{Pauli} is, under certain conditions, of no minor importance. Examples cover systems within a broad range of energy and length scales and pertain to such fundamental questions as  the stability of matter, the structure of atoms and nuclei or the existence of thermodynamic limit \cite{Lieb,LL1972,Bohr}.  This provides a strong motivation for the study of simple fermionic systems whose fundamental properties are induced by the Pauli principle \cite{Thomas2002,GRJin2003,Salomon2003,Jochim2003b,Ketterle2003b,IKS2007,Zwerger2008,GPS2008}.  

One interesting situation arises when the fermions become fully spin—polarized, i.e., {\kb{all particles}} are in the same spin state. The exclusion principle is then manifested entirely in the position space, and the corresponding many—body wave function is necessarily antisymmetric in the spatial coordinates of the particles. Under such circumstances, the Pauli principle has a direct effect on the actual interactions in the system. For instance, the short—range interaction of fermions is at low temperatures determined by the p—wave scattering processes, in contrast to the case of mixtures composed of fermions in different spin states which are governed by the much stronger s—wave interactions \cite{MC2023}.


In this work, we analyze the thermodynamics of a model of interacting fully polarized \emph{dilute} Fermi gas, which we propose on the basis of a Hartree—Fock type microscopic calculation. When the diluteness assumption is exploited, the resulting model is fairly simple and allows for the finite temperature treatment, which appears to have been less explored in the existing literature.  In particular, the resulting equation of state agrees in the zero temperature limit with recent rigorous results obtained for the ground state of the repulsive gas \cite{LauSe}. 

The model incorporates both the short—range repulsion of the p—wave type and the long—range attraction, and the interplay between the two interactions leads to the emergence of a first—order transition between a highly degenerate liquid phase and a gas phase. While the resulting thermodynamics bears certain resemblance to the standard van der Waals theory, it also displays remarkable differences in comparison to it. In particular, the phase transition turns out to exist only for strong enough attractions and, more importantly, it is sensitive to the embedding dimension of the gas. This latter aspect points at the crucial role played by the Pauli principle, as the dimensionality dependence is rooted in the different behavior of the Fermi pressure as function of the density in different dimensions. 
These aspects make the model interesting from the point of view of the theory of phase transitions and critical phenomena, while the analysis might be of relevance for a variety of physical systems where spin polarization is present, ranging from  adequately prepared cold Fermi gases \cite{Blum2004,GuarRadz2007,SGPeng2019,YosUeda2015,Zhang2015,YosUeda2016,Hu2016,JianZhou2018,Zhang2019,MakEns2023} to modelling the interior of neutron stars exposed to strong magnetic fields \cite{Dexheimer2018,Ta2023, Kh2022} \hl{and similar problems arising in nuclear matter \cite{Dexheimer2018} to which phenomenological van der Waals equations of state modified by Fermi statistics have also been applied \cite{Vovchenko2015}.}

The article is organized in the following way. Sec. \ref{mod} contains the necessary definitions and the microscopic derivation of the approximate energy levels and the Helmholtz free energy. The latter is further closely studied in Sec. \ref{thermo}, where \kr{we} discuss the resulting thermodynamics of the model, beginning with a short analysis of the purely repulsive case and subsequently the study of the phase transition which results if the attraction is included, wherein the cases of $d=2$ and $d=3$ are subject to a separate treatment. The results are then summarized in Sec. \ref{last}. The main text is accompanied by an Appendix where we include some calculations relevant for the two--dimensional problem. 

\section{The model}\label{mod} 
\subsection{Preliminaries}
We consider a one--component, fully polarized, interacting Fermi gas composed of $N$ fermions of mass $m$ enclosed in a box of volume $V$  in $d$ dimensions, with $d=2,3$. As mentioned, {\emph{fully polarized}} means here that all fermions occupy the same spin state, which in particular means that one can ignore spin altogether in their description. We assume \kb{that} the fermions interact with each other via an isotropic two--body potential $v$ which generically can be written \kr{as} 
\begin{equation}
\label{poten}
v(x)=v_s(x)+\gamma^{d} v_l(\gamma x) , 
\end{equation}
where $v_s$ is a \kr{compact-support function modelling} the short--range repulsion \kr{of} the particles, while $v_l$ is integrable and models the attraction at longer distances. $\gamma$ is a suitable scaling parameter, and the particular choice of \kr{scaling in Eq.\eqref{poten}} is known as the Kac model \cite{HL1976}, wherein $\gamma$ is sent to zero, corresponding to a very weak and long--ranged interaction. \hl{The Kac scaling ensures that the  $\gamma$-independent integral 
\begin{equation}
\label{kac}
-a:=\int \gamma^{d} v_l(\gamma x) d^dx < 0
\end{equation}
remains finite in the limit $\gamma \rightarrow 0$. Its absolute value is denoted by $a$, which we call the \emph{Kac parameter}. }The  negative sign of the above integral reflects the interparticle attraction at long distances. The choice of the Kac scaling leads to a substantial simplification of the treatment of \kr{long-range attraction} while still producing non trivial effects, as we shall see.

 The fact that the fermions are assumed to be fully polarized affects here mostly the short--range repulsive interactions. The polarization imposes the antisymmetry condition on the positional part of the wave function and precludes the s-wave scattering processes. The effects of the short-range interparticle repulsion are quantified by a single parameter called the p–wave or odd wave scattering length, which we denote by $b$ and which is conveniently defined as \hl{\cite{LauSe}
\begin{eqnarray}
\label{var_sl}
c_{d}\,b^d  := \inf_{\psi} \left\lbrace \int \left(\frac{\hbar^2}{m}|\nabla \psi|^2 +v_s|\psi|^2\right)|x|^2 d^d x \right\rbrace\,,
\end{eqnarray}
where the infimum is taken with respect to functions $\psi(x)$ such  that $\lim_{|x|\rightarrow \infty} \psi(x) = 1 $. }The coefficients $c_d$ are defined as 
\begin{equation}
\label{def::cd}
c_d=
\begin{cases}
& \frac{12\pi \hbar^2}{m}, \quad d=3 \\
&\frac{4\pi \hbar^2}{m}, \quad  d=2.
\end{cases}
\end{equation}
\hl{This variational definition is equivalent to the standard one using the zero--energy scattering equation \cite{Boronat}, and is very useful in the present context.} For instance, in analogy to the Born approximation for the s--wave scattering length \cite{Lan2}, one finds 
\begin{equation}
\label{born3}
c_{d} \,b^d \leq \int v_s(x) |x|^2 d^dx .
\end{equation}
We note a slight abuse in the notation when denoting the two-- and three--dimensional p--wave scattering lengths by the same symbol $b$; this should not lead to any confusion in the remainder of the analysis. 

The parameters $a$ and $b$ thus encode certain information contained in the long--range attractive and short--range repulsive parts, respectively, of a generic interparticle potential $v$, \kj{according to the splitting in Eq.\eqref{poten}}.  

Below we perform a Hartree--Fock calculation whose result is that in the dilute limit \kn{the parameters}  $a$ and $b$ account for the interaction effects completely. \hl{In particular, since the short--range part of $v$ is taken to be purely repulsive in our model, we do not cover effects such as p--wave Cooper pairing of fermions \cite{GuarRadz2007,GRA2005}}.


\subsection{Approximation of the energy levels and the free energy}
Equipped with the representation of both parts of the interaction by means of $a$ and $b$, we now introduce the approximate expression for of the energy levels of the system which we then employ in the \kb{statistical mechanical} analysis. We rely on the following observation: since the p--wave interaction effects are weak, \kn{the interacting gas can be treated as consisting of  free fermions} occupying their one--particle energy levels whose exact form is modified by the p--wave interactions. Accordingly, we identify the microstate of the system with the set of the \kn{single-particle} energy-level occupancies.  The energy of any such microstate can be found by computing the expected value of the hamiltonian $H$ on the appropriate plane--wave Slater determinants $\Psi$, i.e. by performing the Hartree--Fock approximation.  Note that $\Psi$ is characterized completely by the occupation numbers 
$\lbrace n_k \rbrace$, where $n_k=1$ if the mode $k$ is occupied and zero otherwise. 
The Hartree--Fock energy of the Slater determinant \kb{$\Psi$} constructed from $N$ orthonormal functions $|j\rangle$ \kb{equals} 
\begin{equation}
\begin{split}
\langle \Psi| H|\Psi\rangle &= \sum_j n_j \langle j |\frac{-\hbar^2}{2m}\Delta| j\rangle\\
&+ \frac{1}{2}\sum_{j\neq k} n_j n_k\left(\langle jk|v|jk\rangle-\langle jk |v|kj\rangle\right) ,
\end{split}
\end{equation}
where $\frac{-\hbar^2}{2m}\Delta$ stands for the kinetic energy operator and $v$ is the interparticle interaction potential energy in Eq.\eqref{poten}. We employ the periodic boundary conditions and  the plane--wave basis $|k\rangle=\frac{e^{ikx}}{\sqrt{V}}$. Then one obtains  
\begin{equation}
\label{mfs}
\langle jk|v|jk\rangle=\frac{1}{V}\int v(x) d^dx=\frac{1}{V}\left(-a+\int v_s(x)d^dx\right) ,
\end{equation}
where we have employed Eq.\eqref{kac}. The above \emph{direct} term leads to the mean-field expression for the energy. However, it is the \emph{exchange} term discussed below that induces the energy renormalization beyond the above mean--field shift. First, we observe that in the Kac limit the contribution from the attractive part of the interaction to the exchange term vanishes  
\begin{equation}
\begin{split}
\langle jk |v_l |kj\rangle&=\frac{1}{V^2}\int d^dx d^dy\, \gamma^d v_l(\gamma (x-y)) e^{i(k-j)(x-y)}\\ &= \frac{1}{V} \hat{v_l}((k-j)/\gamma)\xrightarrow[\gamma \rightarrow 0]{} 0 
\end{split}
\end{equation}
because $k\neq j$ and the Fourier transform $\hat{v_l}(k)$ vanishes at infinity. 
In the case of the repulsion part \kb{$v_{s}$} the situation is \kb{different. In order to} proceed with our approximation scheme, we use the fact that the potential $v_{s}$ is short--ranged and assume that the gas is dilute and at low temperatures. Accordingly, the typical momenta contributing to the integral can be estimated to lie in the range set by the Fermi energy $\sim n^{1/d}$,  where $n=N/V$,  while the interparticle distances are at most of the order of  $R$, the range of the potential $v_s$, and thus  $|k-j||x|\sim n^{1/d}R\ll 1$. One may thus expand the exponential factor to second order (which actually is one of the crucial steps behind the rigorous bounds developed in \cite{LauSe}) and obtain 
\begin{equation*}
\begin{split}
&\sum_{j\neq k}n_j n_k \langle jk |v_s |kj\rangle \\
&=\sum_{j\neq k} n_j n_k \frac{1}{V^2}\int \kb{d^dx d^dy} \gamma^d v_s(x-y) e^{i(k-j)(x-y)} \\ 
&\approx \sum_{j\neq k}n_j n_k \frac{1}{V}\int \kb{d^dx} v_s(x)\left(1-\frac{((k-j)\cdot x)^2}{2}\right)\\
&=\sum_{j\neq k} \frac{n_jn_k}{V}\left(\int v_s(x)\,\kb{d^dx}  - \frac{(k-j)^2}{2d}\int x^2 v_s(x) \kb{d^dx}\right), 
 \end{split}
\end{equation*}
where in the last step we used the rotation invariance of $v_{s}$. Introducing the total momentum of the Slater determinant $P(\lbrace n_k \rbrace)\equiv \sum_k \hbar k n_k$ and denoting $\tilde{b}=\int x^2 v_s(x) \kb{d^dx}$, one obtains 
\begin{equation}\label{ppsi}
\begin{split}
\langle \Psi| H|\Psi\rangle &= \sum_k \frac{\hbar^2\,k^2}{2m}n_k + \frac{1}{2V}\,\frac{ \tilde{b}}{2d}\sum_{j,k} n_k n_j  (k^2+j^2)\\ 
&-\frac{a}{2V}\sum_{j\neq k} n_k n_j  - \frac{\tilde{b}}{2dV\hbar^2}\,P(\lbrace n_k \rbrace)^2.
\end{split}
\end{equation}
This can be further simplified as follows. First, it can be shown that the term involving the total momentum of the system in Eq.{\eqref{ppsi} (the last term on the rhs of Eq.\eqref{ppsi}) vanishes in the thermodynamic limit. Second, the upper bounds Eq.\eqref{born3} show that the \kr{integral 
$\tilde{b}$} is related to the p-wave  scattering length 
\cite{Blum2004,SGPeng2019,YosUeda2015,Zhang2015,YosUeda2016,Hu2016,Zhang2019} in an analogous way as the integral of the potential is related to the s-wave scattering length. Thus, in the spirit of the original Bogoliubov calculation of the spectrum of the interacting Bose gas \cite{Lan2}, we make here the replacement  \kr{$\tilde{b} \rightarrow c_d b^d$}, with $c_d$ defined in Eq.\eqref{def::cd}. \hl{This is essentially the Born approximation, which is known to provide necessary cancellations for the Bogoliubov approximation to work in the bosonic case \cite{Boccato}.} We find 
the expression for the system energy corresponding to a given microstate $\lbrace n_k\rbrace$ in the following form 
\begin{equation}
\label{eo}
E(\lbrace n_k\rbrace)=\sum_k \frac{\hbar^2 k^2}{2m}\left(1+ \frac{N}{V} B\right) n_k - \frac{a}{2V}N(N-1) ,
\end{equation}
where $N=\sum_{j}n_{j}$, and 
\begin{equation}
\label{BB}
B=\frac{2m}{\hbar^2}\frac{c_d b^d}{2d}=
\begin{cases}
4\pi b^3, \quad d=3 \\
2\pi b^2, \quad d=2\,.
\end{cases}
\end{equation}
\kb{In order to simplify the notation we suppress the subscript $d$ and denote by $B$ a quantity which depends on $d$ and takes different values for two- and three-dimensional systems, \kr{Eq.(\ref{BB}).}  This, however, does not lead to any ambiguity in the analysis and conclusions. }

\kb{In particular, the expression in Eq.\eqref{eo} evaluated} for a microstate corresponding to the Fermi ball agrees with the leading order correction to the ground state energy of a suitably dilute polarized Fermi gas in the thermodynamic limit as developed in \cite{LauSe}. 

For the subsequent thermodynamic analysis, we need to evaluate the expression for the Helmholtz free energy $F(T,V,N)$. The 
canonical partition function \hl{
\begin{equation} 
\label{partfun}
Z(T,V,N)=\sum_{\lbrace n_k\rbrace, \sum_k n_k=N} e^{-\beta E(\lbrace n_k\rbrace)} 
\end{equation}}
evaluated for the system energy given in Eq.(\ref{eo}) straightforwardly leads to the following form of the Helmholtz free energy $F(T,V,N)$ 
\begin{equation}\label{hmhz}
F(T,V,N) = (1+nB)F^{id}\left(\frac{T}{1+nB},V,N\right)-\frac{aN^2}{2V} \,,
\end{equation}
where $F^{\rm{id}}(T,V,N)$ denotes the Helmholtz free energy of the corresponding ideal Fermi gas (i.e., consisting of non-interacting particles of the same mass as in our model) and $n=N/V$. Expression Eq.\eqref{hmhz} forms the starting point of thermodynamic analysis. 

\section{Thermodynamics}
\label{thermo}
\subsection{Purely repulsive gas}
We start with a \kb{brief analysis of the purely repulsive case $a=0$.}
It follows from Eq.\eqref{hmhz} that the equation of state takes the form 
\begin{equation}\label{ps_r}
p(T,n;B)=\left(1+\frac{d+2}{2}n B\right)p^{\rm{id}}\left(\frac{T}{1+nB},n \right)\,,
\end{equation}
where $p^{\rm{id}}(T,n)$ is the pressure of the corresponding ideal Fermi gas \kb{as function of temperature and density, see Eq. \eqref{pmuid} below. When deriving Eq.\eqref{ps_r} the relation $U^{\rm{id}}(T,n)=\frac{d}{2}Vp^{\rm{id}}(T,n)$ was used.}
\kb{This simple result shows that apart from the multiplication by the prefactor $\left(1+\frac{d+2}{2}n B\right)$ the pressure is essentially that of the free Fermi gas evaluated at }the \emph{reduced temperature}
\begin{equation*}
\bar{T}=\frac{T}{1+n B}\,.
\end{equation*}
\kr{On physical grounds, because of the interparticle repulsion,} one expects $p(T,n)\geq p^{\rm{id}}(T,n)$. This is not easily seen directly from Eq.\eqref{ps_r} because on one hand the prefactor $\left(1+\frac{d+2}{2}n B\right)$  increases the pressure, but on the other hand the reduced temperature $\bar{T}$ is lower than the actual one, thus lowering the pressure at a given density. It is of interest to see how the overall increase \kb{of pressure} can be deduced from Eq. \eqref{ps_r}. \kb{This also serves as a check of consistency of our model with the thermodynamic formalism. }
Let us compute
\begin{equation*}
\left(\frac{\partial p}{\partial B}\right)_{T,n}=n\left(Dp^{\rm{id}}(\bar{T},n)-\frac{1+D nB}{1+nB}\bar{T}\left(\frac{\partial p^{\rm{id}}(\bar{T},n)}{\partial \bar{T}}\right)_{n}\right)\,,
\end{equation*}
where 
\begin{equation}\label{D}
D=\frac{d+2}{2}.
\end{equation}
With the help of thermodynamic identity 
$\left(\frac{\partial U}{\partial V}\right)_{T,N} = -p + T\left(\frac{\partial p}{\partial T}\right)_n$, where $U$ is the internal energy,  the above relation can be rewritten in the following form 
\begin{eqnarray}
\label{pn}
\left(\frac{\partial p}{\partial B}\right)_{T,n} = \frac{D-1}{1+nB}\left(n^2B D\left[n\left(\frac{\partial p^{\rm{id}}}{\partial n}\right)_{\bar{T}} - p\right] \right. \nonumber \\ 
\left. + \, n^2\,(1+ nB + n^2DB^2)\, \left(\frac{\partial p^{\rm{id}}}{\partial n}\right)_{\bar{T}}  \right).
\end{eqnarray}
\kb{It follows from Eq.(\ref{pn}) that $\left(\frac{\partial p}{\partial B}\right)_{T,n} >0$. Indeed,}  the pressure is an increasing function of the density at constant temperature, and thus $p^{\rm{id}}(\bar{T},n)$ is an increasing function of the density at constant $\bar{T}$. Thus the last term in Eq.\eqref{pn} is positive, while one can verify by direct computation using Eq. \eqref{pmuid} below that the first term is also positive. This, together with $p(T,n)|_{B=0}=p^{\rm{id}}(T,n)$, shows $p(T,n)\geq p^{\rm{id}}(T,n)$, as expected. 
We plot the resulting isotherms for different values of $B$ in Fig. \ref{pres_r}.
Incidentally, the above argument also verifies that the isothermal compressibility \kb{is positive.  This follows from }
\begin{equation}
\left(\frac{\partial p}{\partial n}\right)_{T}=\left(\frac{\partial p}{\partial n}\right)_{\bar{T}}+\frac{B}{n}\left(\frac{\partial p}{\partial B}\right)_{T,n}\,.
\end{equation} 
\begin{figure}[!htb]
\includegraphics{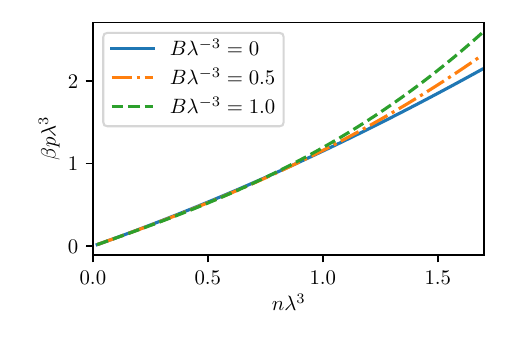}
\caption{Isotherms for the purely repulsive polarized Fermi gas ($a=0$), at different values of the parameter $B \lambda^{-3}$.}
\label{pres_r}
\end{figure}

\subsection{Equation of state and the chemical potential equation}
\hl{In the previous discussion, we verified that is thermodynamically consistent to treat the repulsive polarized Fermi gas at finite temperatures by means of the ground state correction and the effective reduced temperature $T\rightarrow T/(1+nB)$. In what follows, we investigate this system with  the attraction term included. 
The (canonical) pressure reads then simply
\begin{equation}\label{presque}
p(T,n)=\left(1+\frac{d+2}{2}n B\right)p^{\rm{id}}\left(\frac{T}{1+nB},n \right)-a\frac{n^2}{2}.
\end{equation}
This form is similar to the classical van der Waals theory, with the hard--core contribution replaced by the combined effect of the Fermi pressure and the short--range repulsion quantified by parameter $B$.  In Fig. \ref{iso}, we plot the isotherms corresponding to $p(T,n)$ and observe that they do not satisfy the stability condition $\left(\frac{\partial p}{\partial n}\right)_{T}>0$ for sufficiently large $a$ and low $T$, which, just like in the classical van der Waals theory, marks the existence of a phase transition between high and low density phases in the system. The order parameter here is the difference in the bulk densities. We note that, at least far away from the critical temperature, the coexisting densities are such that the system changes is behavior from essentially classical to highly degenerate at the transition point. This is evident when the relevant fragments of the isotherms are compared with the high and low degeneracy asymptotics of the pressure in Fig. \ref{iso}. This points at the possibility of observing exotic effects pertinent to highly degenerate Fermi systems in the high density phase. }
\begin{figure}[!htb]
\includegraphics{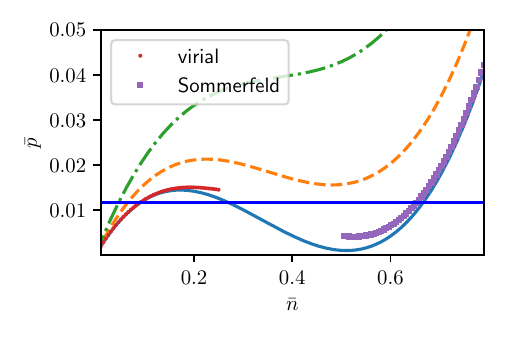}
\caption{\hl{The isotherms of the three-dimensional attractive polarized Fermi gas in the canonical ensemble. The dimensionless pressure $\bar{p}$ and density $\bar{n}$ are defined in the text, see Eq. \eqref{def01}. The curves correspond to three distinct temperatures, with the solid line being the lowest and the dot--dashed line the highest among them, at a fixed value of $a=1.3 a_c$, with $a_c$ defined in Eq. \eqref{def01}. At low temperatures, the isotherms experience negative compressibility, marking the onset of a first-order phase transition. The horizontal solid line marks the transition pressure for the lowest isotherm, calculated from the Maxwell construction known from elementary thermodynamics \cite{Callen}. The distance between the extremal points of intersection of this line with the isotherm determine the difference in the bulk densities at the transition point. The pressure at the low density phase is well approximated by the virial expansion for the almost--classical Fermi gas while the pressure at the high density phase is close to the one following the Sommerfeld expansion for the highly degenerate gas \cite{Huang}. The corresponding approximations are plotted in the relevant regions in comparison to the lowest isotherm. }}
\label{iso}
\end{figure}

\hl{Although the equation of state is very similar in appearance to the standard van der Waals theory, we wish to investigate whether the details of the transition, as quantified e.g. by the values of the critical parameters, reveal differences in comparison to the standard case stemming from the degeneracy. To overcome the lack of stability and describe the transition within the proper thermodynamic formalism}, we work in the grand canonical formalism by fixing  the chemical potential $\mu$ and temperature $T$. The actual equilibrium pressure $p(T,\mu)$ is then obtained via the Legendre transform 
\begin{equation}
-p(T,\mu)= \inf_{n\geq 0} \left(f(T,n)-\mu n\right) ,
\end{equation}
where $f(T,n) = F(T,V,N)/V$ is the Helmholtz free energy density. With the help of  the relation 
$f(T,n)=-p(T,n)+n \mu(T,n) $ 
the Legendre transform takes the following form 
\begin{equation}
\label{lt}
\begin{split}
-p(T,\mu)&= \inf_{n\geq 0} \left\lbrace-(1+nB)p^{\rm{id}}\left(\frac{T}{1+nB},n\right)\right.\\
&\left.+(1+Bn)n\mu^{\rm{id}}\left(\frac{T}{1+nB},n\right)-\frac{an^2}{2}-\mu n\right\rbrace .
\end{split}
\end{equation}
\kb{The ideal Fermi gas functions} are given implicitly by 
\begin{equation}
\label{pmuid}
\begin{split}
n\lambda^d = &f_{\frac{d}{2}}(e^{\beta\mu^{\rm{id}}(T,n)}), \\
p^{\rm{id}}(T,n) = &\frac{k_B T}{\lambda^d}f_{\frac{d}{2}+1}(e^{\beta\mu^{\rm{id}}(T,n)}), 
\end{split}
\end{equation} 
where $k_B$ is the Boltzmann's constant, $\lambda=\sqrt{\frac{2\pi\hbar^2}{mk_B T}}$ is the thermal de Broglie wavelength, and 
\begin{equation}
\kb{f_j(e^z):= - \sum_{l=1}^{\infty}\frac{(-e^{z})^l}{l^j}} =-\mathrm{Li}_{j}(-e^{z})
\end{equation}
is the Fermi function. 
In order to perform the Legendre transform and find the equation of state, we compute the derivative of the function in the brackets in Eq.\eqref{lt} with respect to $n$ and set it equal to zero. Keeping in mind that the $n$ dependence is also present in the first argument of both $p^{\rm{id}}$ and $\mu^{\rm{id}}$ and using Eq.\eqref{pmuid} together with the identity 
\begin{equation}
\frac{d f_j(-e^z)}{d z}=f_{j-1}(-e^z)
\end{equation}
we obtain the \emph{chemical potential equation}
\begin{eqnarray}
\label{mu_eq}
\mu = (1+nB)\mu^{id}\left(\frac{T}{1+nB},n\right) + \nonumber \\ 
+ (D-1)\,B p^{id}\left(\frac{T}{1+nB},n\right) - a\,n,
\end{eqnarray}
\kb{whose solutions $n(T,\mu)$ are the main object of our further discussion.}

In the remainder of the article, we analyze the case $a>0$
more closely, focusing on low temperatures and covering the cases of $d=3$ and $d=2$ separately.

\subsection{Fully polarized Fermi gas in d = 3}
To gain some insight we first consider $T=0$. In this case the chemical potential equation Eq.\eqref{mu_eq} in $3D$ reduces to 
\begin{equation}\label{mu_eqT0d3}
\mu=(1+nB)\varepsilon_F+\frac{3}{5}Bn \varepsilon_F - an \equiv \varphi_{0}^{(3)}(n),
\end{equation}
where $\varepsilon_F=C_3 n^{2/3}$ is the Fermi energy, with $C_3=\frac{\hbar^2}{2m}(6\pi^2)^{2/3}$.
The $T=0$ equation has three solutions $n(T,\mu)$ as long as the derivative of $\varphi_{0}^{(3)}(n)$ is negative in a finite interval of the densities. \kb{This corresponds to two-phase coexistence  and  a first-order transition between them.} The existence of such an interval is possible only if the interparticle attraction \kb{represented by parameter $a$} is strong enough. The critical point beyond which only one solution exists is given by
\kb{\begin{equation}
\frac{\partial \varphi_{0}^{(3)}}{\partial n} =0 \,\,\,,\,\,\, \frac{\partial^2 \varphi_{0}^{(3)}}{\partial n^2} =0.
\end{equation}
The above equations give the critical density
\begin{equation}
\label{critden}
n_c=\frac{1}{8 B}
\end{equation}
and  the critical value of the Kac parameter $a_c$ 
\begin{equation}\label{ac_v}
a_c=2 C_3 B^{1/3}=\frac{\hbar^2}{m} \left(6\pi^2 B \right)^{1/3}
\end{equation}}
The first-order transition exists only for $a > a_{c}$. \hl{The critical value $a_c$ is such that the typical attractive interaction between a pair of fermions localized in volume $B$, equal to $a/B$ in the Kac scaling, and evaluated at $a=a_{c}$, is of the order of their kinetic energy $\sim \frac{\hbar^2}{m B^{2/3}}$, which is necessarily non--zero due to the Pauli principle. This gives $a_c\sim \frac{\hbar^2}{m}B^{1/3}$ as in \eqref{ac_v}. Note that no such bound is present in the classical van der Waals equation of state with hard core particles, where the transition takes place for arbitrary values of $a$, regardless of the radii of the particles. 
}

The critical value of the chemical potential at $T=0$ is
\begin{equation}
\mu_c=\frac{C_3}{20 B^{2/3}}=\frac{\hbar^2}{40m} \left(\frac{6\pi^2}{B}\right)^{2/3}>0.
\end{equation}
We also note that the function $\varphi_{0}^{(3)}(n)$ is bounded from below, meaning that solutions to Eq.\eqref{mu_eqT0d3} exist only for large enough values of the chemical potential. This is not surprising, as the chemical potential of the ideal gas at $T=0$ is  positive. Non--trivial solutions of Eq.\eqref{mu_eqT0d3} appear for $\mu\leq 0$ only for  $a\geq a_m=1.134 a_c$.  This in turn imposes the constraint $a_c<a<a_m=1.134 a_c$ for the phase transition to occur at zero temperature. For $a > a_m$, only the high density phase exists as a stable entity, with $n>\frac{1}{2B}$ . 
We note that if the critical density $n_{c}$ at $T=0$, Eq.\eqref{critden}, is expressed \kb{via} the $p-wave$ scattering length \kb{$b$} one obtains $n_{c}b^3=0.00994$ which means that the gas is clearly dilute in these circumstances. This is consistent with one of our initial assumptions leading to the derivation of the model. 

\kb{We now turn to the case $T>0$} and introduce the dimensionless variables 
\begin{equation}
\label{def01}
\begin{split}
\bar{n}=nB, &\quad t=\frac{\pi}{\sqrt{12}}\frac{k_B T B^{2/3}}{C_3}, \,\,\,\quad \bar{\mu}=\frac{\mu B^{2/3}}{C_3},\\
 &\quad \bar{a}=\frac{a }{2C_3 B^{1/3}} = \frac{a}{a_{c}}, \,\,\,\bar{p}= \frac{p B^{5/3}}{C_{3}}.
 \end{split}
\end{equation}
The chemical potential equation Eq.\eqref{mu_eq} can be solved numerically. For $\bar{a}>1$ the equation admits three distinct roots, provided that $t$ is small enough and \kr{$\bar{\mu}$} lies in a suitable range. The equilibrium density is given by the root that yields the the smallest value potential term in the brackets in Eq. \eqref{lt}, which in turn yields the equilibrium pressure. 

\kr{Fig.\ref{iso_3d}} displays the isotherms $\bar{p}(t,\bar{\mu})$ of the three-dimensional fully polarized Fermi gas corresponding to different values of parameter $a$.  In particular, the $T=0$ isotherms are also included in these plots and they illustrate the discussion presented above. The behavior of the isotherms depends on the value of parameter $\bar{a}$. In general, the qualitative behavior of the isotherms differs for small  $\bar{a}$ and for large $\bar{a}$. In particular, \kr{for} $\bar{a} < \kb{1}$ the isotherms are smooth while for $\bar{a} >\kb{1}$ one observes lines of first-order phase transitions terminating at the critical points. 

\begin{figure}[!htb]
\includegraphics{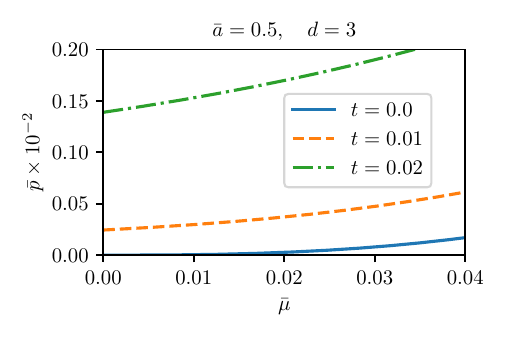}
\includegraphics{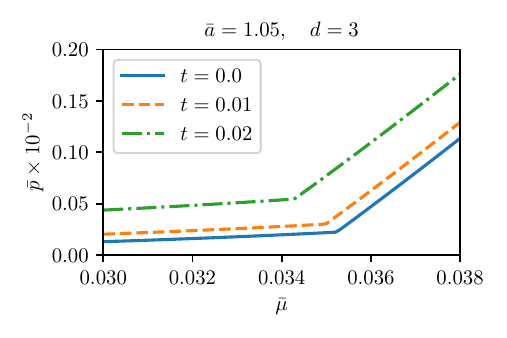}
\includegraphics{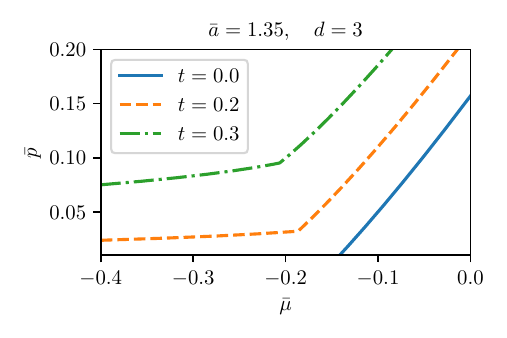}
\caption{The isotherms \hl{$\bar{p}(t,\bar{\mu})$} of the fully polarized Fermi gas in $d=3$ for different values of the parameter $\bar{a}$. For $\bar{a}<1$  the isotherms are smooth and no transition appears. For  $\bar{a}>1$ the isotherms display a discontinuity of their derivatives \hl{at} low temperatures, marking a first--order transition which is present also at $T=0$, as both values of the derivatives are non--vanishing at the coexistence point. For high enough values of $\bar{a}$, the density of the gaseous phase tends to zero as $T\rightarrow 0$. }
\label{iso_3d}
\end{figure} 
 
 The characteristics of the critical points can be determined analytically in the low temperature domain. We consider the regime 
\kb{$\bar{a} >1$ with $\bar{a}$ close to the boundary value 1.}  Then, the phase transition is present at \kb{non-zero temperatures with very low critical temperature}. In this regime, the density \kb{difference} between the coexisting phases \kb{is  small. The densities of both coexisting phases} are of the order of $n_c$, and the gas can be considered dilute, in agreement with our assumptions. A further advantage of restricting analysis to this regime  is that we can effectively treat the problem analytically by virtue of the well--known Sommerfeld expansion applied to the pressure and chemical potential of the Fermi gas: 
\begin{equation}
\label{smfd}
\begin{split}
& \bar{\mu}^{\rm{id}}(t,\bar{n})\approx  \bar{n}^{2/3}\left(1-\frac{t^2}{\bar{n}^{4/3}}\right),\\
&\bar{p}^{\rm{id}}(t,\bar{n})\approx \frac{2}{5} \bar{n}^{5/3}\left(1+5 \frac{t^2}{\bar{n}^{4/3}}\right),
\end{split}
\end{equation}
and thus \kb{the chemical potential equation, Eq.\eqref{mu_eq}, reads}
\begin{equation}
\label{mu_eq_smfd}
\begin{split}
\bar{\mu}\approx & (1+\bar{n})\bar{n}^{2/3}\left(1-\frac{t^2}{\bar{n}^{4/3}}\right)\\
&+\frac{3}{5}\bar{n}^{5/3}\left(1+\frac{5 t^2}{ \bar{n}^{4/3}}\right)-2\,\bar{a}\,\bar{n}\,\equiv \varphi^{(3)}(t,n)\,.
\end{split}
\end{equation}
We are first interested in determining the critical point at which both the first and second derivative \kb{of the function 
$\varphi^{(3)}(t,\bar{n})$ 
 with respect to $\bar{n}$ vanish, see Eq.\eqref{mu_eq_smfd}. In the regime $\bar{a}\approx 1$ one expects that} the critical density is very close to the zero--temperature critical density $\bar{n}_c=\frac{1}{8}$ and that the critical temperature is close to zero.
Thus, we expand the right--hand side of Eq.\eqref{mu_eq_smfd} to \kb{the} third order in $\bar{n}-\bar{n}_{c}$ (this is the minimal order which yields three different solutions to Eq.\eqref{mu_eq_smfd}) and then solve the resulting equations treating $\bar{a}-1$, 
$\bar{n}-\bar{n}_{c}$ and $t$ as small parameters. In this way one obtains 
\kb{
\begin{equation}
\label{3dasm}
\begin{split}
&t_c\approx \frac{9}{16}\sqrt{\frac{3}{11}}\left(\bar{a}-1\right)^{\frac{1}{2}}\\
&\bar{n}_c\approx \frac{1}{8}+\frac{21}{22}(\bar{a}-1)\\
&\bar{\mu}_c\approx \frac{1}{20}-\frac{5}{11}(\bar{a}-1)\,.
\end{split}
\end{equation}}
These equations determine the critical point in the regime \kb{$\bar{a}\approx 1$,} see Fig. \ref{crit_t}. \hl{Note that $t_c$ vanishes as $a$ approaches $a_c$ while $n_c$ tends to $1/(8B)$, in accordance with the zero temperature calculation.
Importantly, we note the power law behavior of $t_c\sim \sqrt{a-a_c}$ and $\bar{n}_c$ in $d=3$, in contrast with the classical van der Waals gas where the critical temperature is linear in $a$.} Moreover, in the next section we shall show that \kb{in the case of 
$t_{c}$ the power law behavior does not hold in $d=2$. This difference is rooted in the different scaling of the Fermi energy with the density in different dimensionalities, and marks the sensitivity of the transition to the Pauli pressure.}


\subsection{Fully polarized Fermi gas in d=2}

In order to discuss the two-dimensional systems we introduce the dimensionless variables 
\begin{equation}
t=\frac{k_B T B}{a_0}, \quad \bar{n}=n B, \quad \bar{a}=\frac{a}{a_0}, \quad \bar{\mu}=\frac{\mu B}{a_0}, 
\end{equation}
where  $a_0=\frac{2\pi\hbar^2}{m}$. \kb{At this point we note that the dimensionless parameters $\bar{a}$ are defined in $d=2$ and $d=3$ cases in such a way  that their critical value is 1 both for $d=2$ and $d=3$. In the  $T=0$ case the chemical potential equation takes the simple form}
\begin{equation}
\label{muT0}
\bar{\mu}=\frac{3}{2}\bar{n}^2+(1-\bar{a})\bar{n}.
\end{equation}
\kb{For $\bar{a} \leq 1$ only one solution $\bar{n}(\bar{\mu})$ exists and corresponds to $\bar{\mu}>0$.} For $\bar{a}>1$ two solutions exist for  a certain range of negative $\bar{\mu}$ values but only one of them corresponds to a stable minimum of the expression in the curly brackets on the right hand side of Eq.(\ref{lt}) evaluated at $T=0$.  Hence, in the $d=2$ case there is no phase transition \kb{at $T=0$}, in contrast to the  $d=3$ case. 
\begin{figure}[!htb]
\includegraphics{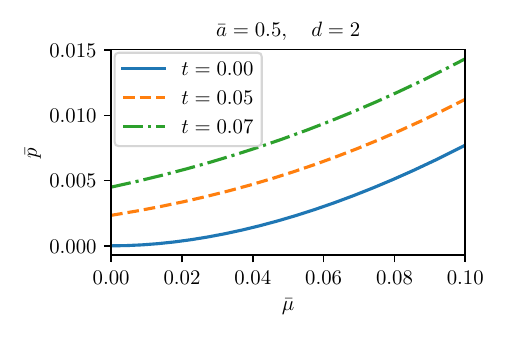}
\includegraphics{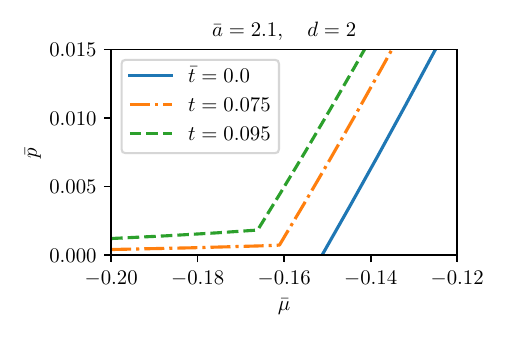}
\caption{The isotherms \kr{$\bar{p}(t,\bar{\mu})$} of the two-dimensional fully polarized Fermi gas for subcritical (upper \kr{panel}) and supercritical (lower \kr{panel}) values of parameter $\bar{a}$. The first--order transition is present if the coupling is large enough $\bar{a}>1$, as marked here by the discontinuity of the derivative of the pressure with respect to the chemical potential in the supercritical case. Note that the gaseous phase density becomes zero at $T=0$ for all values of parameter $\bar{a}$ for which the transition is present for $T>0$. This zero-temperature behavior remains in contrast to the case $d=3$ where there is a window of parameter $\bar{a}$ for which the transition survives down to $T=0$.  }
\label{iso_2d}
\end{figure}
In the $T>0$ case Eq.\eqref{mu_eq} reads
\begin{eqnarray}
\label{mu_eq2d}
\bar{\mu} = t \,\ln \left(\exp{\left(\frac{\bar{n}(1+\bar{n})}{t}\right)} -1\right) \,\,\,\,\,\,\, \\
+\frac{t^2}{(1+\bar{n})^2}f_2\left(\exp{\left(\frac{\bar{n}(1+\bar{n})}{t}\right)}-1\right)-\bar{a}\bar{n} \equiv \varphi^{(2)}(t, \bar{n})\,. 
\nonumber
\end{eqnarray}
The solutions of this equation represent the equilibrium density as a function of $t$ and $\bar{\mu}$. In Fig. \ref{iso_2d} we plot the resulting isotherms for different temperatures; for comparison with the $d=3$ case see Fig. \ref{iso_3d}. We observe the existence of a first--order transition  provided $\bar{a}>1$ and the temperature is small enough. \hl{Similarly as in $3D$ and in contrast to the classical van der Waals theory, we observe that a high enough attraction is needed for the transition to occur. However, in contrast to the $d=3$ case, the critical value of $a$ does not depend on $B$, as it drops out when the kinetic energy $\sim \frac{\hbar^2}{m B}$ is equated with the Kac attraction of a pair of particles enclosed in a $2D$ volume $B$, $a/B$. Another difference with the $3D$ gas is that the density of the gaseous phase tends to zero as $t\rightarrow 0$, regardless of the value of parameter $\bar{a}$. }\kb{This is in} accordance with the discussion of $t=0$ case above. In fact, in $d=3$ the gaseous phase density attains a non--zero value at $T=0$ provided  $1<\bar{a}<1.134$. Thus one observes a qualitative difference in behavior of two- and three-dimensional systems, sumarized in Fig. \ref{pd} where we present the coexistence lines in the $T,\mu$ space in the two dimensionalities, for different values of $a$. 
\begin{figure}[!htb]
\includegraphics{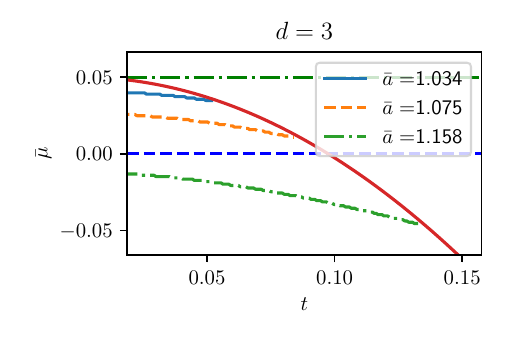}
\includegraphics{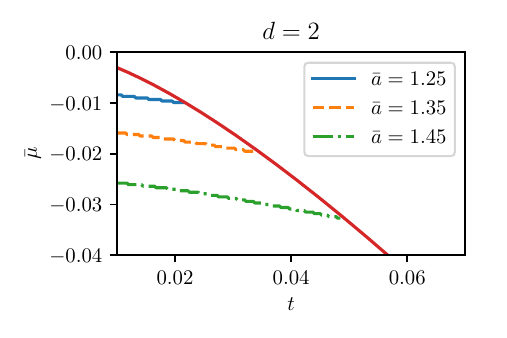}
\caption{\hl{The coexistence lines in the $\bar{\mu},t$ plane, separating the low density (low $\bar{\mu}$) and large density (large $\bar{\mu}$)  phases, in dimension $d=3$ (upper panel) and $d=2$ (lower panel), for different values of the attraction parameter $a$.  The dimensionless variables 
$\bar{\mu}$ and $t$ are defined in Eq. \eqref{def01}.  As $a$ decreases, the lines get shorter, and the critical temperature approaches zero as $a\rightarrow a_c$.  The transition is absent above the dot--dashed line for all temperatures, while below the dashed line it is absent for $T=0$; this window closes completely for $d=2$.  The solid lines are the asymptotic envelopes of the line of critical points corresponding to different values of $a$, see Eq. \eqref{3dasm}.}}
\label{pd}
\end{figure}

This different behavior of two- and three-dimensional systems is also seen in the asymptotic behavior of the \kb{critical temperatures and densities expressed as functions of corresponding parameter $\bar{a}$ close to its minimal value $\bar{a}=1$.} From the technical point of view the distinction between $d=2$ and $d=3$ cases is seen already at the level the Sommerfeld expansion. In $d=2$ it does \emph{not} provide the correct asymptotic behavior because it neglects the exponential corrections which turn out to be \kb{relevant} and are responsible for the appearance of the low--density solution. \kb{In order to find the critical point we equate to zero the first- and second-order derivatives of the function $\varphi^{(2)}(t,\bar{n})$ in Eq.\eqref{mu_eq2d} and solve} the resulting equations in which we keep the exponential factors and the polynomial terms in  leading order in $\bar{n}$ and $t$. The details of this calculation are provided in the Appendix.  The result is
\begin{equation}
\label{lambert}
\begin{split}
&t_c \approx - \frac{\bar{a}-1}{3W_{-1}(\frac{1-\bar{a}}{e})} \approx - \frac{\bar{a}-1}{3\ln(\bar{a}-1)} \\
&\bar{n}_c \approx \frac{\bar{a}-1}{3} +\frac{\bar{a}-1}{3\ln(\bar{a}-1)}  \\
\end{split}
\end{equation}
where $W_{-1}(z)$ denotes the lower branch of the Lambert function \cite{Lambert}, i.e. the negative valued solution of $we^w=z$ for small $z$, behaving asymptotically as $\ln(- z)$ for small $|z|$. 
In Fig. \ref{crit_t}, we plot the values of the critical temperature found numerically together with the asymptotic formulae Eq.\eqref{lambert}. \hl{We observe that $n_c$ approaches zero as $\bar{a}$ reaches its critical value, in accordance with the lack of the transition at $T=0$, where the low density phase reduces to vacuum. Remarkably, the asymptotic form of the expression for the critical temperature does not reduce to a simple power law characterizing three-dimensional systems, Eq. \eqref{3dasm}, still less the classical van der Waals critical temperature $T_c\sim a$.} This highlights the sensitivity of the system's properties to the embedding dimensionality, rooted in different scaling laws of the Fermi pressure in different dimensions and, together with the complete absence of the transition at $T=0$ in $d=2$, marks an intriguing departure from the standard van der Waals theory \kb{which is due} to the Pauli principle. \\
\begin{figure}[!htb]
\includegraphics{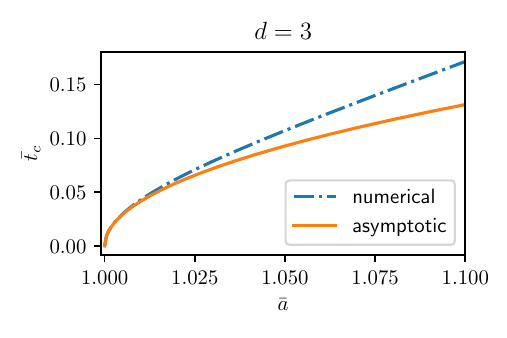}
\includegraphics{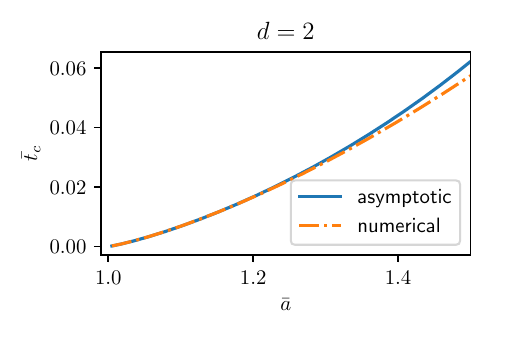}
\caption{The critical temperature in $d=3$ (upper plots) and $d=2$ (lower plots) as a function of the coupling $\bar{a}$, together with the asymptotic laws Eq.\eqref{lambert} and Eq.\eqref{3dasm}, respectively, as a function of the coupling parameter $\bar{a}$. While in $d=3$ the critical temperature close to the critical coupling follows a simple power law, the $d=2$ case is \kb{ different, with the asymptotic law involving the logarithmic correction}.  }
\label{crit_t}
\end{figure}

\section{Summary}\label{last}

We have constructed and analyzed a simple statistical mechanical model of a fully polarized dilute Fermi gas with short--range repulsion and long--range attraction. It  relies on the introduction of suitably modified one--particle energy levels whose form agrees with recently obtained bounds on the ground state energy of the repulsive gas and is here heuristically justified via a Hartree--Fock type calculation. The repulsion and the attraction are quantified by two parameters:  the p--wave scattering length $b$, and the integral of the attractive part of the potential $-a$. The derived form of the approximate system's energy  enables us to find the Helmholtz free energy, which is then employed to find the equation of state via the Legendre transform.  Our analysis shows that the system undergoes a first--order phase transition in dimensions $d=3$ and $d=2$, provided the parameter $a$ is large enough. We \kb{evaluate the critical temperature and density for  $a$ close to the minimal value necessary for the transition to occur. We find} a remarkable quantitative difference in their behavior as function of $a$ \kb{corresponding} to the $d=3$- and $d=2$-systems. In addition, the transition is absent at $T=0$ in $d=2$ for the entire range of $b$ and $a$ values. On the contrary, in $d=3$, there exists a window of  parameter $a$-values for which the transition also takes place in the ground state. These discrepancies between the $d=3$ and $d=2$ cases  are ultimately rooted in the different scaling relations connecting the Fermi pressure and the density in these dimensionalities. \kb{In this way, the evidence of a nontrivial phase behavior of  the polarized Fermi gas has been provided within a study which goes beyond the usual van der Waals theory.  Our analysis suggests that one can expect to observe simple manifestations of the Pauli principle at the macroscopic level in similar systems. }

\emph{Acknowledgements.} We thank Asbjørn Bækgaard Lauritsen for helpful discussions. Financial support from the National Science Center of Poland (NCN) via grant 2020/37/B/ST2/00486 (K.M.) and \kn{2021/43/B/ST3/01223} (M.N.) is gratefully acknowledged.

\newpage
\appendix*
\widetext{
\section{}
Here we sketch the derivation of the  asymptotic laws Eq.\eqref{lambert} governing the behavior of the critical temperature and \kb{critical} density \kb{of two-dimensional system} in the vicinity of the minimal value of parameter $\bar{a}$. This minimal value $\bar{a}=1$ corresponds to the onset of the phase transition.  We start with the chemical potential equation $\bar{\mu} = \varphi^{(2)}(t,\bar{n})$, Eq.{\eqref{mu_eq2d}. }
The critical point parameters $t_c(\bar{a}),\bar{n}_c(\bar{a})$ solve the equations below
\begin{equation}\label{der::def}
\begin{split}
0=\frac{\partial \varphi^{(2)} }{\partial \bar{n}}&=-\bar{a}+\frac{(1+2\bar{n})^2}{1+\bar{n}}\frac{\exp{\left(\frac{\bar{n}(1+\bar{n})}{t}\right)}}{\left(\exp{\left(\frac{\bar{n}(1+\bar{n})}{t}\right)}-1\right)}
-\frac{2t^2}{(1+\bar{n})^3}f_2\left(e^{\left(\frac{\bar{n}(1+\bar{n})}{t}\right)}-1\right),\\
0= \frac{\partial^2 \varphi^{(2)}}{\partial \bar{n}^2}&=\frac{(1+2\bar{n})}{t(1+\bar{n})^2}\frac{e^{\left(\frac{\bar{n}(1+\bar{n})}{t}\right)}}{\left(e^{\left(\frac{\bar{n}(1+\bar{n})}{t}\right)}-1\right)^2
}\left(\left.-1-\right.\right. 
 \left. 5\bar{n}-8\bar{n}^2-4\bar{n}^3+3t\left(e^{\left(\frac{\bar{n}(1+\bar{n})}{t}\right)}-1\right)\right)+
 \frac{6t^2}{(1+\bar{n})^4}f_2\left(e^{\left(\frac{\bar{n}(1+\bar{n})}{t}\right)}-1\right).
\end{split}
\end{equation}
We are interested in the behavior of $t_c $ and $\bar{n}_c$ in the regime $\bar{a}\approx 1$, where the coexistence line is very short and accordingly $t_c$ is small. From the absence of the transition at $t=0$ we conclude that also $n_c$ is close to zero. Accordingly, we expand the terms in Eq.\eqref{der::def} that are polynomial or rational functions in \kn{variables} $t$ and 
$\bar{n}$ to leading order. On the other hand, since a priori we do not know the behavior of $\bar{n}_c/t_c$  
we need to keep terms like $e^{\bar{n}_{c}(1+\bar{n}_{c})/t_{c}}$ in the calculations. In this way the second equation above takes the following form
\begin{equation}
\label{A12}
0 \approx \frac{e^{\frac{\bar{n}(1+\bar{n})}{t}}}{\left(e^{\frac{\bar{n}(1+\bar{n})}{t}} -1\right)^2}\,
\left(-1-7\bar{n}\right) + 3t\left(e^{\left(\frac{\bar{n}(1+\bar{n})}{t}\right)}-1\right)+
 \frac{6t^3}{(1+\bar{n})^2}\,f_2\left(e^{\left(\frac{\bar{n}(1+\bar{n})}{t}\right)}-1\right).
\end{equation}
We consider the asymptotic regime $\bar{a}\searrow1$  in which $t_{c}$, $\bar{n}_{c}$ and their ratio $t_{c}/\bar{n}_c$ tend to zero. In this regime Eq.(\ref{A12}) simplifies to
\begin{equation} 
\label{pre-w}
3\,t_{c}\, - \,e^{- \frac{\bar{n}_{c}(1+\bar{n}_{c})}{t_{c}}}\left(1+ 3t_{c} +7\bar{n}_{c}\right) = 0 ,
\end{equation}
where $f_2(e^z)\approx \frac{z^2}{2}$ for large $z$ has been used. 
\kb{We use this relation to get rid of the exponential terms in the upper equation in Eq.\eqref{der::def} and obtain
\begin{equation}
\bar{n}_c \approx \frac{\bar{a} -1}{3} - t_{c}. 
\end{equation}
When inserted back into Eq.\eqref{pre-w}, it gives 
\begin{equation}
e^{\frac{\bar{a}+2}{3}\,\left(\frac{\bar{a} -1}{3t_{c}}-1\right)}  - 1=\frac{1 + \frac{7}{3}(\bar{a}-1)-7t_{c}}{3t_{c}}.
\end{equation}
Thus, in the asymptotic regime $\bar{a}\searrow1$ one obtains \cite{Lambert, NIST} 
\begin{equation}
\label{lambert2}
\begin{split}
&t_c \approx - \frac{\bar{a}-1}{3W_{-1}(\frac{1-\bar{a}}{e})} \approx - \frac{\bar{a}-1}{3\ln(\bar{a}-1)} \\
&\bar{n}_c \approx \frac{\bar{a}-1}{3} +\frac{\bar{a}-1}{3\ln(\bar{a}-1)} , \\
\end{split}
\end{equation}} where we have chosen the lower branch of the Lambert W function because $1-\bar{a}$ is small and negative and $e\frac{\bar{a} -1}{3t_{c}}$ is large, and used the asymptotic behavior of the Lambert function $W_{-1}(x)\approx \ln(-x)$ for $x$ small and negative.

\end{document}